\documentclass[aps,prd,twocolumn,groupaddress,tighten,nofootinbib]{revtex4}
\usepackage{amssymb}
\usepackage{amsmath,color}
\usepackage{epsfig}
\usepackage{graphicx,epsf,epsfig}
\usepackage{bbm}
\begin{document}
\title{Non-unitarity effects in a realistic low-scale seesaw model}
\author{Michal Malinsk\'y}
\email{malinsky@kth.se}

\author{Tommy Ohlsson}
\email{tommy@theophys.kth.se}

\author{He Zhang}
\email{zhanghe@kth.se}

\affiliation{Department of Theoretical Physics, School of
Engineering Sciences, Royal Institute of Technology (KTH) --
AlbaNova University Center, Roslagstullsbacken 21, 106 91 Stockholm,
Sweden}
\begin{abstract}
We analyze the structure of the non-unitary leptonic mixing matrix
in the inverse seesaw model with heavy singlets accessible at the
LHC. In this model, unlike in the usual TeV seesaw scenarios, the
low-scale right-handed neutrinos do not suffer from naturalness
issues. Underlying correlations among various parameters governing
the non-unitarity effects are established, which leads to a
considerable  improvement of the generic non-unitarity bounds. In
view of this, we study the discovery potential of the non-unitarity
effects at future experiments, focusing on the sensitivity limits at
a neutrino factory.
\end{abstract}
\maketitle
\section{Introduction} \label{sec:intro}

One of the most intriguing open questions in particle physics
nowadays is the origin of the unprecedentedly small neutrino masses
and the peculiar leptonic mixing parameters observed in neutrino
oscillation experiments. If neutrinos are Majorana particles, the
seesaw approach
\cite{Minkowski:1977sc,Yanagida:1979as,Mohapatra:1979ia,Schechter:1980gr,Lazarides:1980nt,Mohapatra:1980yp,Foot:1988aq}
provides a natural link between the lepton sector observables and
the dynamics underlying the breakdown of the lepton number.  In the
simplest schemes, the relevant scale typically falls into the
$10^{11}~{\rm GeV}$ to $10^{14}~{\rm GeV}$ range, and thus seems
never accessible to direct tests. However, there are options for a
low-scale seesaw model, and in principle, the corresponding features
of new physics can be probed in the forthcoming accelerator
experiments such as those at the Large Hadron Collider (LHC).

The prospects of testing the origin of neutrino masses at colliders
are determined by the scale of the underlying physics and one would
appreciate it not to be far above the electroweak scale. If this is
the case, a plethora of new effects such as the non-unitarity
\cite{Antusch:2006vwa} of the Maki--Nakagava--Sakata (MNS) leptonic
mixing matrix (imprinted into specific patterns of non-standard
neutrino interactions, the enhancement of the lepton flavor
violation (LFV) phenomena, etc.) can be within the reach of near
future experimental facilities. This, in turn, provides a
complementary strategy for unveiling the seesaw structure.

Unfortunately, within the popular type-I seesaw framework, this
strategy is plagued by naturalness issues. Indeed, the requirement
of reproducing the sub-eV neutrino masses tends to be incompatible
with visible non-unitarity effects. The reason is the simplistic
structure of the type-I seesaw formula, that in a convenient
notation reads $m_{\nu}=F M F^{T}$. Here $M$ is the right-handed
(RH) neutrino mass scale and $F=M_{D}M^{-1}$ (with $M_{D}$ denoting
the Dirac neutrino mass matrix) corresponds to the structure
governing the non-unitarity effects. One option of reconciling the
TeV-scale $M$ with the sub-eV light neutrino masses $m_\nu$ is to
take $F$ to be of the order of $10^{-5}$ which, however, leads
neither to any appreciable non-unitarity effects nor to LHC signals.
Alternatively, one can invoke a cancellation in the matrix structure
of the seesaw formula
\cite{Buchmuller:1991tu,Pilaftsis:1991ug,Ingelman:1993ve,Heusch:1993qu,Kersten:2007vk},
which is also not natural unless extra assumptions are made about
the flavor structure of the model.

Concerning the other simple seesaw schemes, the situation in the
type-III case is essentially identical to that in the type-I case,
namely, there is no significant non-unitarity effect without
fine-tuning the matrix structure \cite{Bajc:2006ia,Bajc:2007zf}. On
the other hand, the type-II scheme with a light Higgs triplet offers
distinctive features at the colliders as well as high-precision
neutrino experiments and has been studied in great detail in
e.g.~Refs.~\cite{Perez:2008ha,Malinsky:2008qn} and references
therein. However, the leptonic mixing matrix is exactly unitary at
the renormalizable level, since there is no RH sector the light
states could admix with. Thus, none of the simple seesaw
realizations of the Majorana neutrino masses provides a satisfactory
framework accommodating both the collider phenomenology and the
non-unitarity effects at an experimentally accessible level.

In this work, we therefore focus on the simplest inverse seesaw
model \cite{Mohapatra:1986bd}, which shares all the virtues of the
type-I seesaw scenario. In particular, it has the same predictive
power concerning the non-unitarity as well as collider effects, yet
providing a completely natural description of the sub-eV light
neutrino masses. The key point is that in this framework the $B-L$
breaking mass insertion in the seesaw formula is decoupled from the
RH neutrino mass scale, and thus, the light neutrino masses do not
impose any stringent bounds on the size of the $F$ parameters
governing the interesting phenomenology. In this respect, the
inverse seesaw scenario can be regarded as the simplest natural
scheme accommodating RH neutrinos accessible to LHC whilst admitting
complementary tests exploiting the would-be non-unitarity of the
leptonic mixing matrix.

This work is organized as follows: In Sec.~\ref{sec:model}, we
comment in more detail on the structure of the inverse seesaw model
and compare its parameter space to the conventional type-I seesaw
scheme. In Sec.~\ref{sec:formalism}, we focus on the emergence and
basic implications of the non-unitarity effects. The characteristic
correlations between various non-unitarity effects in the context of
some of the future experiments, in particular the neutrino factory
and LHC direct searches, are studied in detail in
Sec.~\ref{sec:phenomenology}.

\section{The inverse seesaw model} \label{sec:model}
The inverse seesaw model \cite{Mohapatra:1986bd} is an extension of
the type-I seesaw scenario with three extra Standard Model (SM)
gauge singlets $S^{\alpha}$ coupled to the RH neutrinos
$\nu_{R}^{\alpha}$ through the lepton number conserving couplings of
the type $\overline{\nu_{R}^{c}}S$, while the traditional RH
neutrino Majorana mass term is forbidden by extra symmetries. It is
thus only through a dimensionful parameter $\mu$ in the
self-coupling $\mu\overline{S}{S^{c}}$ the lepton number is broken
and one can arrange $\mu$ to be arbitrarily small in a technically
natural manner. The $9\times 9$ mass matrix in the
$\{\nu_{L},\nu_{R}^{c}, S^{c}\}$ basis then reads
\begin{eqnarray}
M_{\nu}=\left(\begin{array}{ccc}
0 & M_{\rm D} & 0 \\
M^T_{\rm D} & 0 &M_{\rm R} \\
0 & M_{\rm R}^{T} & \mu
\end{array}\right) \ ,\label{eq:M}
\end{eqnarray}
where $M_{D}$ and $M_{R}$ are generic $3\times 3$ complex matrices
representing the Dirac mass terms in the $\nu_{L}$-$\nu_{R}$ and
$\nu_{R}$-$S$ sectors.
\footnote{Similar mass matrices can be obtained in some technicolor
models \cite{Appelquist:2002me,Appelquist:2003hn}.}
Without loss of generality, one can always choose a basis in which
$\mu$ is real and diagonal: $\mu={\rm diag} (\mu_1,\mu_2,\mu_3)$.
The mass matrix $M_{\nu}$ can be diagonalized by means of a $9\times
9$ unitary transformation
\begin{eqnarray}\label{Vmatrix}
V^{\dagger}M_{\nu}V^{*}= \bar{M}_\nu = {\rm
diag}(m_i,{M}_{j}^{n},{M}_{k}^{\tilde{n}}) \
\end{eqnarray}
{with ($i,j,k=1,2,3$), where $m_i$ denote the masses of the
left-handed neutrinos, while the RH neutrinos $\nu_{R}^{\alpha}$ and
the extra singlets $S^{\alpha}$ are almost maximally admixed into
three pairs of heavy Majorana neutrinos $(n_{j},\tilde n_{j})$.
Since $n_{j}$ and $\tilde n_{j}$ have opposite CP parities and
essentially identical masses $M_{j}^{n}$ and $M_{k}^{\tilde{n}}$
(with a splitting of the order of $\mu$), they can be regarded as
components of three pseudo-Dirac neutrinos. Assuming $\mu \ll M_{\rm
D} < M_{\rm R}$, the light neutrino Majorana mass term is  given
approximately by
\begin{eqnarray}\label{eq:mnu}
m_{\nu}\simeq F\mu F^{T} \,,
\end{eqnarray}
where, as in the type-I case, $F\equiv M_{\rm D} (M_{\rm
R}^{T})^{-1}$. Notice that the structure of the neutrino mass matrix
is essentially identical to the type-I formula, i.e., it depends on
two basic building blocks -- the flavor structure of the $B-L$
breaking mass insertion (the RH neutrino Majorana mass matrix $M$ in
type-I and $\mu$ in the inverse seesaw setting) and the ratio $F$.
Since in both cases, the LFV, the non-unitarity effects as well as
the LHC rates are driven only by $F$ and the spectra of the heavy
components involved in the charged currents, the relevant parameter
spaces of these two scenarios are equivalent. From this perspective,
the inverse seesaw enjoys a similar predictivity as the type-I case,
but in a more realistic (yet experimentally interesting) regime.

\section{The non-unitarity effects} \label{sec:formalism}
The light neutrino mass matrix can be diagonalized by a unitary
transformation $U$
\begin{eqnarray}\label{eq:mnuV}
U^\dagger m_{\nu} U^* = \bar{m}_\nu \
\end{eqnarray}
with $\bar{m}_\nu = {\rm diag} (m_1,m_2,m_3)$. In the standard
(i.e., CKM-like) parametrization one has
\begin{eqnarray}\label{eq:parametrization}
U  = P_{\rho}R_{23}P_{\delta}R_{13}P_{\delta}^{-1}R_{12}P_{M} \ ,
\end{eqnarray}
where $R_{ij}$ correspond to the elementary rotations in the
$ij=23$, $13$, and $12$ planes (parametrized in what follows by
three mixing angles $c^{}_{ij} \equiv \cos \theta^{}_{ij}$ and
$s^{}_{ij} \equiv \sin \theta^{}_{ij}$), $P_{\delta}={\rm
diag}(1,1,{\rm e}^{{\rm i}\delta})$, and $P_{M}={\rm diag}({\rm
e}^{{\rm i}\alpha_{1}/2}, {\rm e}^{{\rm i}\alpha_{2}/2},1)$ contain
the Dirac and Majorana CP phases, respectively. The $P_{\rho}={\rm
diag}({\rm e}^{{\rm i}\rho_{1}}, {\rm e}^{{\rm i}\rho_{2}}, {\rm
e}^{{\rm i}\rho_{3}})$ phases entering the charged currents are
usually rotated away in the SM context (or in the regime the RH
sector decouples) but must be kept in the current scenario. Even in
the basis where the charged-lepton mass matrix is diagonal, $U$ is
only a part of the mixing matrix governing neutrino oscillations.
Instead, one should look at the upper-left sub-block of the full
$9\times 9$ matrix $V$ in Eq.~\eqref{Vmatrix}
\begin{eqnarray}
V=\left(\begin{array}{cc}
V_{3\times3} & V_{3\times6} \\
V_{6\times3} & V_{6\times6}
\end{array} \right) \ .
\end{eqnarray}
For $M_{\rm R}$ not far above the electroweak scale and a reasonably
small $\mu$, it is sufficient to consider the form of $V$ at the
leading order in $F$.\footnote{As will be shown in
Sec.~\ref{sec:phenomenology}, the experimental constraints indicate
that such an approximation is reasonably good.}
The full (non-unitary) MNS mixing matrix then reads
\cite{Kanaya:1980cw,Schechter:1981cv} (in  the notation of
Ref.~\cite{Altarelli:2008yr}):
\begin{eqnarray}\label{N33}
N\equiv V_{3\times 3} \simeq \left(1-\frac{1}{2}FF^{\dagger}\right)U \ ,
\end{eqnarray}
and the $3\times 6 $ block participating in the charged currents
reads $K\equiv V_{3\times 6} \simeq \left(0,F\right)
V_{6\times6}\;.$ These structures control all the observables of our
further interest. The defining flavor eigenstates
$\{\nu_{L},\nu_{R}^{c}, S^{c}\}$ correspond to superpositions of the
mass eigenstates $\{\hat\nu_{L},n,\tilde n\}$, and the left-handed
neutrinos entering the electroweak currents obey $\nu_{L} \simeq N
\hat \nu_{L}+K P\ ,$ where
$P=(n_{1},..,n_{3},\tilde{n}_{1},..,\tilde{n}_{3})$. Using
Eqs.~\eqref{eq:mnu} and \eqref{eq:mnuV}, one can write
$F=U\sqrt{\bar{m}_\nu} O \sqrt{\mu^{-1}} $ with $O$ being a complex
orthogonal matrix \cite{Casas:2001sr}. At the leading order in the
non-unitarity of $N$, the entries  of the unitary matrix $U$ can be
parametrized by the measured values of the leptonic mixing
parameters, and thus, $F$ depends only on $m_{1}$, $O$, and $\mu$.
As a simple example,\footnote{Note that this proposition is, indeed,
in the spirit of no extra fine-tuning in the neutrino sector.
Nevertheless, in what follows, we will consider more general
settings as well.} for $O=\mathbbm{1}$, one obtains
\begin{eqnarray} \label{D}
\bar{m}_{\nu}=D \mu D^{T}\; ,
\end{eqnarray}
where $D={\rm diag} (d_1,d_2,d_3)$ is a real and diagonal matrix
acting as a {\it compensator} between the entries of
$\mu$ and $\bar{m}_\nu$.

The charged current Lagrangian in the mass basis is
\begin{eqnarray}
{\cal L}_{\rm CC} = -\frac{g}{\sqrt{2}}   \overline{\ell_{\rm L}}
\gamma^\mu \left( N \hat \nu_{L} + KP\right)W^-_\mu + {\rm H.c.}
\end{eqnarray}
This has several important implications:

i) The conventional unitary leptonic mixing matrix is replaced by a
non-unitary matrix $N$. One can rewrite Eq.~\eqref{N33} as
\begin{eqnarray} \label{eq:NF}
N  = U \left(1-\frac{1}{2} \sqrt{\bar{m}_\nu} O \mu^{-1} O^\dagger
\sqrt{\bar{m}_\nu} \right) \ ,
\end{eqnarray}
which means the inverse seesaw structure yields a characteristic
pattern of the unitarity violation. The elements of $U$ are merely
rescaled and the non-unitarity effects exhibit correlations with
non-trivial experimental implications. Moreover, Eq.~\eqref{eq:NF}
justifies the validity of choosing the diagonal basis for charged
leptons.

We will use a variant of a convenient parametrization advocated in
Ref.~\cite{Altarelli:2008yr}, namely $N=(1-\eta)U$, which is
particularly suitable for the studies of neutrino oscillation
phenomena. The small Hermitian matrix $\eta$ in the setting of our
interest obeys
\begin{eqnarray} \label{eq:eta}
2\eta=FF^\dagger =U \sqrt{\bar{m}} O  \mu^{-1} O^\dagger
\sqrt{\bar{m}} U^\dagger \ .
\end{eqnarray}
It is clear that the phases $P_{\rho}$ do not affect the magnitudes
but only the overall phases of the individual entries of $\eta$.

ii) The heavy neutrinos $n_{i}$ and $\tilde{n}_{i}$ couple to the
gauge sector of the SM, and thus, if kinematically accessible, can
be produced at hadron colliders. Due to their pseudo-Dirac nature,
the lepton number violating collider signatures will be suppressed
with respect to the fine-tuned type-I and III scenarios, where the
heavy states are Majorana particles. Therefore, if $K \propto F$ is
sizeable and the RH sector is accessible, one could expect
appreciable rates in the LFV channels.

iii) The LFV decays $\ell^-_\alpha \rightarrow \ell^-_\beta \gamma$
are controlled by the magnitude of $F$ that, similar to the
fine-tuned type-I case, is not suppressed by the light neutrino
masses, and thus can be remarkable \cite{Deppisch:2004fa}.

In what follows, we will comment in more detail on the three points
above, and focus, in particular, on the discovery potential for the
non-unitary effects of the future neutrino oscillation experiments.

\section{Phenomenological consequences} \label{sec:phenomenology}
\subsection{Constraints on non-unitarity parameters}
\label{sec:constraints} \vspace{-2.5mm} In general, the deviation of
the leptonic mixing matrix from unitarity is constrained namely from
the universality tests of the weak interactions, rare leptonic
decays, invisible width of the $Z$-boson, and neutrino oscillation
data. The current 90~\% C.L. bounds on the entries of $\eta$ are summarized in
Refs.~\cite{Antusch:2008tz,Antusch:2009pm}: $|\eta_{ee}| < 2.0
\times 10^{-3}$, $|\eta_{\mu\mu}| < 8.0 \times 10^{-4}$,
$|\eta_{\tau\tau}| < 2.7 \times 10^{-3}$, $|\eta_{e\mu}|< 3.5 \times
10^{-5}$, $|\eta_{e\tau}|< 8.0 \times 10^{-3}$, and
$|\eta_{\mu\tau}|< 5.1 \times 10^{-3}$.

Concerning the shape of $O$, we will consider three basic
situations. First (case I), let $O$ be a unit matrix. Given the
correlations of the non-unitarity effects, in particular the simple
structure of Eq.~\eqref{D}, the six generic parameters
$|\eta_{\alpha\beta}|$ are no longer independent and one can exploit
Eq.~\eqref{eq:eta} to improve some of these bounds by almost an
order of magnitude. Indeed, Eq.~\eqref{eq:eta} reads in an explicit
matrix form
\begin{equation}
\eta \simeq \! \frac{1}{2}P_{\rho}\!\!\left( \begin{matrix} d^2_1\!
-\!d^2_{12} s^2_{12} & -d^2_{12}s_{12}c_{12}c_{23} &
d^2_{12}s_{12}c_{12}s_{23} \\ . & d_{3}^{2}-d^2_{312} c^2_{23} &
d^2_{312} s_{23} c_{23} \\ . &. & d_{3}^{2}-d^2_{312} s^2_{23}
\end{matrix}\right)\!\!P_{\rho}^{\dagger},\label{eq:etaN}
\end{equation}
where $d^2_{12}\equiv d_{1}^{2}-d_{2}^{2}$, $d^2_{312}\equiv
(d_{3}^{2}-d_{1}^{2})s_{12}^{2}+(d_{3}^{2}-d_{2}^{2})c_{12}^{2}$,
the small $\theta_{13}$ effects have been neglected, and the omitted
entries follow from the Hermitian property of $\eta$. Examining
Eq.~\eqref{eq:etaN}, the correlations induced in the present
framework can be readily obtained, and the current upper bounds on
$|\eta_{\alpha\beta}|$ are upgraded to $|\eta_{e\tau}| < 3.5 \times
10^{-5}$, $|\eta_{\mu\tau}| < 8.0 \times 10^{-4}$, $|\eta_{ee}| <
1.6 \times 10^{-3}$, and $|\eta_{\tau\tau}| < 8.0 \times 10^{-4}$.

Second (case II), let $\mu$ be flavor blind, i.e., $\mu=\mu_0
\mathbbm{1}$, and $O$ arbitrary. Equation (\ref{eq:eta}) then yields
$\eta = \frac{1}{2} \mu_0^{-1} U \sqrt{\bar{m}_\nu} {\rm exp}
(2\,{\rm i}A) \sqrt{\bar{m}_\nu} U^\dagger$, where $A$ is a real
antisymmetric matrix \cite{Pascoli:2003rq}. Following the same
strategy, one obtain improved bounds $|\eta_{e\tau}| < 2.3 \times
10^{-3}$ and $|\eta_{\mu\tau}| < 1.5 \times 10^{-3}$, while the
other experimental limits are saturated.

Third (case III), one can consider the most general setting relaxing
also the degeneracy in the matrix $\mu$. In such a case, all the
current experimental bounds can be saturated simultaneously.
Nevertheless, this in general does not mean that any configuration
of the values of $|\eta_{\alpha\beta}|$ that may be measured in
future experiments can be accommodated. However, a detailed analysis
of this most generic setting is out of the scope of this work and
will be performed elsewhere.

Hence, from the point of view of the future neutrino oscillation
experiments, both case-I and II naturally accommodate ``sizeable''
(i.e., a few per mil) non-unitarity effects in the $\nu_\mu
\rightarrow \nu_\tau$ channel. In principle, this can be used to
test the minimal inverse seesaw model.

\subsection{Sensitivity at a neutrino factory}
\vspace{-2.5mm} For a non-unitary leptonic mixing matrix $N$, the
vacuum neutrino oscillation transition probability $P_{\alpha\beta}$
can be written as \cite{Ohlsson:2008gx}
\begin{eqnarray}\label{eq:P}
P_{\alpha\beta} &=&  \sum_{i,j} {\cal F}^i_{\alpha\beta} {\cal
F}^{j*}_{\alpha\beta} - 4 \sum_{i>j} {\rm Re} ({\cal
F}^i_{\alpha\beta} {\cal F}^{j*}_{\alpha\beta} )\sin^2\!\left(
\frac{\Delta m^{2}_{ij}L}{4E}\right) \nonumber \\ &+& 2
\sum_{i>j}{\rm Im} ( {\cal F}^i_{\alpha\beta} {\cal
F}^{j*}_{\alpha\beta} ) \sin\left(\frac{ \Delta m^{2}_{ij} L}{2
E}\right) \ ,
\end{eqnarray}
where $\Delta m^{2}_{ij} \equiv m^2_i - m^2_j$ are the neutrino
mass-squared differences and ${\cal F}^i$ are defined by
\begin{eqnarray}\label{eq:F}
{\cal F}^i_{\alpha\beta} \equiv \sum_{\gamma ,\rho} ( R^*)_{\alpha
\gamma } ( R^*)^{-1}_{\rho \beta } U^*_{\gamma i} U_{\rho i} \
\end{eqnarray}
with the normalized non-unitary factor
\begin{eqnarray}\label{eq:R}
R_{\alpha\beta} \equiv \frac{(1-\eta)_{\alpha\beta}}
{\left[(1-\eta)(1-\eta^\dagger)\right]_{\alpha\alpha}} \ .
\end{eqnarray}
When Earth matter effects are considered, one can replace the vacuum
quantities $U$ and $m_i$ by their effective matter counterparts, see
e.g.~Ref.~\cite{Meloni:2009ia}.
\begin{figure}[t]
\vspace{-4mm}
\begin{center}
\includegraphics[width=9.3cm]{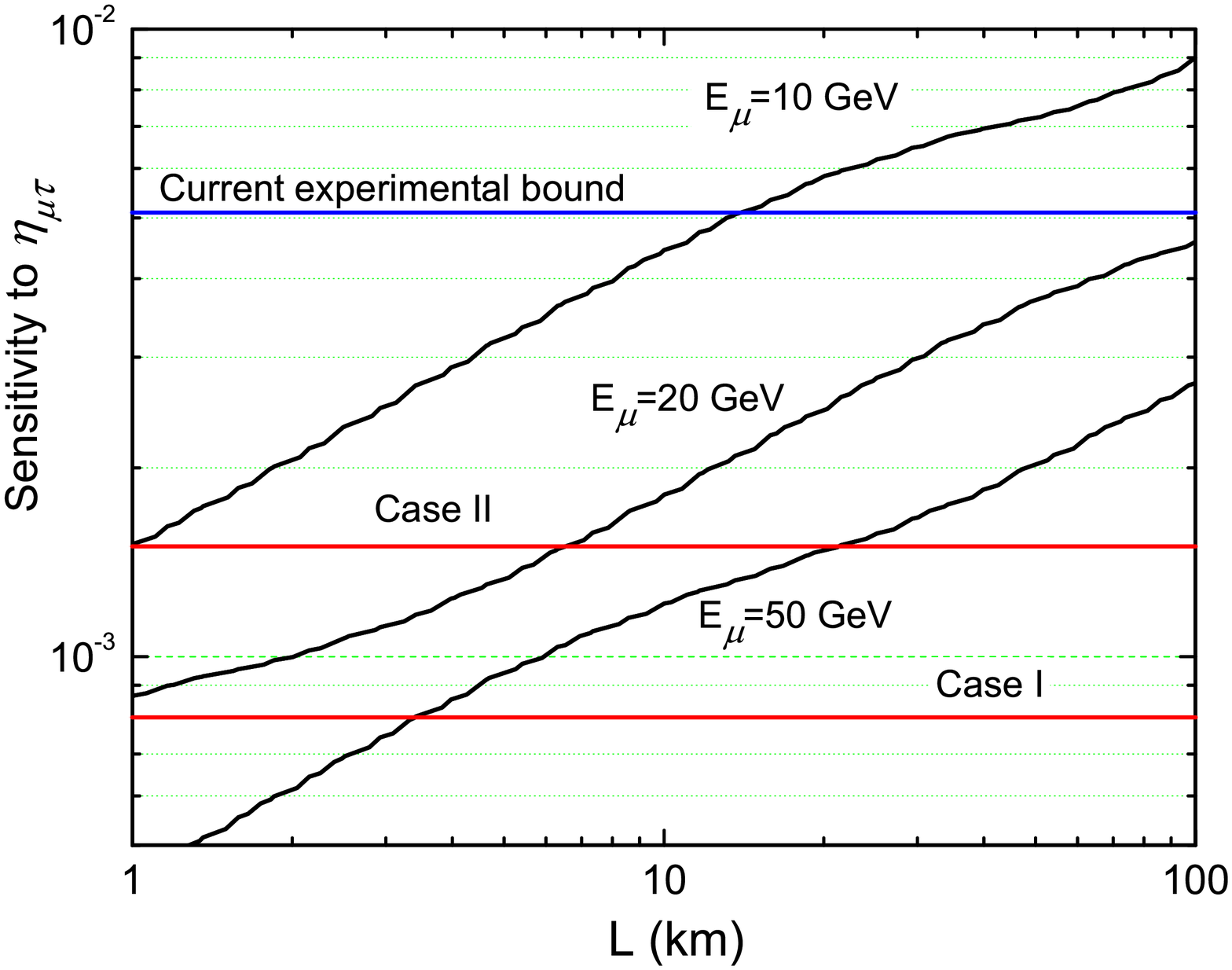}\mbox{}
\end{center}
\vspace{-8mm} \caption{\label{fig:OPERA}Sensitivity limits at
$90~\%$ C.L.~on $\eta_{\mu\tau}$ in the inverse seesaw model as a
function of $L$ corresponding to the first two (mandatory) terms in
Eq.~\eqref{eq:Pmutau}.}  \vspace{-0.2cm}
\end{figure}

As we argued in Sec.~\ref{sec:constraints}, the $\nu_\mu \rightarrow
\nu_\tau$ channel is the most favorable channel to constrain the
model, since it is correlated with $\eta_{\mu\tau}$ which is by far
the largest off-diagonal entry of $\eta$. In this respect, the best
sensitivity is generally provided by short baseline setups, since
the standard oscillation effects are suppressed by sines of $L$ in
such setups
\cite{FernandezMartinez:2007ms,Goswami:2008mi,Donini:2008wz}.

Thus, in what follows, we will consider the transition probability
$P_{\mu\tau}$ for a neutrino factory with a short
enough baseline length. We neglect the matter effects, the tiny
mixing angle $\theta_{13}$, and the small mass-squared difference
$\Delta m^{2}_{21}$. In such a case the transition probability with
non-unitarity effects reads \cite{FernandezMartinez:2007ms}
\begin{eqnarray}\label{eq:Pmutau}
P_{\mu\tau} &\simeq & 4|\eta_{\mu\tau}|^2 + 4s^2_{23}c^2_{23}
\sin^2\left(\frac{\Delta m^{2}_{31} L}{4E}\right)
 \nonumber \\
 &-&  4|\eta_{\mu\tau}|\sin\delta_{\mu\tau}
s_{23}c_{23} \sin\left(\frac{\Delta m^{2}_{31} L}{2E}\right) \ ,
\end{eqnarray}
where the last term is CP odd due to the phase $\delta_{\mu\tau}$ of
$\eta_{\mu\tau}$, and hence induces distinctive CP-violating effects
in neutrino oscillations \cite{Altarelli:2008yr}. Since the model
under consideration does not provide any information about
$\delta_{\mu\tau}$, we will stick to the most pessimistic scenario
with $\delta_{\mu\tau}=0$, and the non-unitarity effects emerge only
from the first ``zero distance'' term in Eq.~\eqref{eq:Pmutau},
which is quadratic in $|\eta_{\mu\tau}|$. For any non-negligible
values of $\sin\delta_{\mu\tau}$ one can then expect the
non-unitarity effects to be even more pronounced, since the CP-odd
contribution is linear in $|\eta_{\mu\tau}|$.

Let us illustrate the feasibility of observing such a signal in a
typical neutrino factory setup with an OPERA-like near detector with
fiducial mass of 5 kt. We assume a setup with approximately $
10^{21}$ useful muon decays and five years of neutrino running. We
make use of the GLoBES package \cite{Huber:2004ka,Huber:2007ji} with
a slight modification of the template Abstract Experiment Definition
Language (AEDL) file for neutrino factory experiments
\cite{Autiero:2003fu,Huber:2006wb}. In Fig.~\ref{fig:OPERA}, we
display the sensitivity to $|\eta_{\mu\tau}|$ as a function of the
baseline length $L$ for the near detector. One can observe that such
a setup provides indeed an excellent probe for this type of
non-unitarity effects. As expected, the sensitivity is decreasing
with the baseline length due to the oscillation effects. Thus, a
distance $L \lesssim 100~ {\rm km}$ would be favorable for the near
detector.
\footnote{Note that, practically, an extremely short baseline setup
(i.e., $L=3~{\rm km}$ and $E_{\mu}=25~{\rm GeV}$) may not be
efficient, since the beam divergence is not comparable with the size
of detector. See Ref.~\cite{Tang:2009na} for detailed discussions.}

\subsection{Potentially interesting LHC signatures}
\vspace{-2.5mm}
\begin{figure}[t]
\begin{center}
\includegraphics[width=5.5cm]{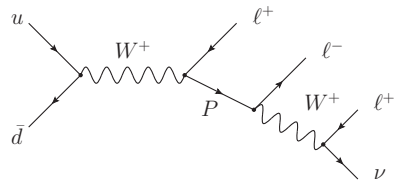}
\caption{\label{fig:LHC} Scheme of the potentially interesting LHC
signature with three charged leptons and missing energy emerging at
appreciable rates in the model under consideration.} \vspace{-0.7cm}
\end{center}
\end{figure}
Since the amount of lepton number violation in the current setting
is small (driven by $\mu$) \cite{Keung:1983uu}, the striking LHC
signature of the fine-tuned type-I and III models with like-sign
leptons in the final state $pp\to
\ell^{\pm}_{\alpha}\ell^{\pm}_{\beta} + \mbox{jets}$
\cite{Datta:1993nm,Han:2006ip,Aguila:2007em} is suppressed.
Technically, the suppression emerges from the interplay between the
graphs with internal lines of the $n$ and $\tilde n$ type that tend
to cancel due to the opposite CP parities of these states leaving
behind only factors proportional to $\mu$. However, the lepton
flavor violating processes are insensitive to this effect and in
principle one can expect observable signals in the channels with
small SM background. For example, one very interesting and
prospective channel is the production of three charged leptons and
missing energy \cite{delAguila:2008hw}, i.e., $pp\to
\ell_{\alpha}^{\pm} \ell_{\beta}^{\pm}\ell_{\gamma}^{\mp}
\nu(\bar\nu)+\text{jets}$, which is depicted in Fig.~\ref{fig:LHC}.
Another possible process is the pair production of charged leptons
with different flavor and zero missing energy, i.e., $pp\to
\ell_{\alpha}^{\pm}\ell_{\beta}^{\mp} + \mbox{jets}$. Note that it
is difficult to make the observation of this channel at the LHC due
to the large SM background \cite{Aguila:2007em}.

\subsection{Lepton flavor violating decays}
\vspace{-2.5mm}
The heavy pseudo-Dirac singlets $P$ entering the charged currents
due to the non-unitarity effects also contribute to the lepton
flavor violating decays $\ell_\alpha \rightarrow \ell_\beta
\gamma$.
The amplitude is proportional to $(FF^{\dagger})_{\alpha\beta}$ that
measures the amount of non-unitarity in the diagonal sub-blocks of
$V$ \cite{Ilakovac:1994kj}. In the standard type-I seesaw scenario
(i.e., without cancellations), one has approximately $FF^{\dagger}=
{\cal O}(m_{\nu}M_{R}^{-1})$, and therefore ${\rm
BR}\left({\ell_\alpha \rightarrow \ell_\beta \gamma}\right) \propto
{\cal O} (m_{\nu}^{2})$ indicates a strong suppression of LFV
decays. However, in the inverse seesaw case, one can have sizeable
$F=M_{D} M_{R}^{-1}$ in spite of $m_{\nu}\to 0$. Thus, appreciable
LFV rates could be obtained even for strictly massless light
neutrinos \cite{Bernabeu:1987gr}.

\section{conclusions} \label{sec:conclusions}
In this letter, we have elaborated on the non-unitarity effects in
neutrino oscillations due to the relative proximity of the
electroweak scale and the scale of the would-be right-handed
neutrinos in the inverse seesaw model of light neutrino masses.
Unlike the traditional type-I and III seesaw scenarios, this
framework does not suffer from naturalness issues even if the heavy
sector is low enough to be accessible at the LHC. Moreover, it can
accommodate sizeable lepton flavor violating effects.

The simplistic flavor structure of the model, which is argued to be
essentially equivalent (in complexity) to the type-I seesaw
scenario, yields distinctive correlations between the
phenomenological parameters $\eta_{\alpha\beta}$ governing the
non-unitarity of the leptonic mixing matrix. In view of the possible
significant off-diagonal entry $\eta_{\mu\tau}$, we have studied the
discovery potential of a neutrino factory experiment with an
OPERA-like near detector in the $\mu\to \tau$ channel and presented
the relevant sensitivity limits to $\eta_{\mu\tau}$. Potentially
interesting signatures at the LHC and in the lepton flavor violating
decays have also been discussed. \vspace{-6mm}
\begin{acknowledgments}
\vspace{-2.5mm} We wish to thank Walter Winter, Mattias Blennow, and
Enrique Fern{\'a}ndez-Mart{\'i}nez for useful comments. The work was
supported by the Royal Swedish Academy of Sciences (KVA) [T.O.], the
G{\"o}ran Gustafsson Foundation [H.Z.], the Royal Institute of
Technology (KTH), contract no.~SII-56510 [M.M.], and the Swedish
Research Council (Vetenskapsr{\aa}det), contract no.~621-2008-4210
[T.O.].
\end{acknowledgments}

\end{document}